\documentclass[12pt]{article}
\usepackage{graphicx}

\begin{document}

\begin{center}
\Large {\bf  Massless Cosmic Strings in Expanding Universe}
\end{center}

\bigskip
\bigskip

\begin{center}
D.V. Fursaev
\end{center}

\bigskip
\bigskip

\begin{center}
{\it Dubna State University \\
     Universitetskaya str. 19\\
     141 980, Dubna, Moscow Region, Russia\\

  and\\

  the Bogoliubov Laboratory of Theoretical Physics\\
  Joint Institute for Nuclear Research\\
  Dubna, Russia\\}
 \medskip
\end{center}

\bigskip
\bigskip

\begin{abstract}
Circular massless cosmic strings which move with the speed of light in the de Sitter universe are described. Construction of 
the background geometry is based  on parabolic isometries of the de Sitter spacetime. Microscopic circular cosmic strings may appear at the Planck epoch and then grow up to
the Hubble size.  We analyze: images of the strings,  influence of strings on trajectories of matter, formation of overdensities, and shifts 
of energies of photons. 
These effects allow one to discriminate massless strings from their massive cousins. The present work extends our results on straight massless
cosmic strings in Minkowsky spacetime to curved backgrounds.
\end{abstract}

\newpage

\section{Introduction}\label{intr}

Massless cosmic strings (MCS) are one-dimensional objects of zero thickness which move with the speed of light. MCS in a flat spacetime can be obtained from 
common massive cosmic strings \cite{Kibble:1976sj} as a limiting case,  when the velocity of the string reaches the speed of light, mass tends to zero, while energy  remains finite \cite{Barrabes:2002hn}. As a result of this limit, a holonomy along a closed countour around a massive string 
is transformed into a non-trivial holonomy around the MCS \cite{vandeMeent:2012gb}. The holonomies of MCS belong to a parabolic  subgroup 
of Lorentz transformations. Therefore, similarly to massive strings MCS allow global gravitational effects. The effects look as mutual transformations
of trajectories of massive bodies or light rays, when the string moves in between two trajectories.

A method how to describe physical effects around massless cosmic strings in a flat spacetime
by using the parabolic transformations was developed in \cite{Fursaev:2017aap}.  The main features of MCS are the following. 

First, massless strings are specified by an energy $E$ per unit length, which, like the energy
of a photon, depends on a frame of reference where it is measured. MCS, as opposed to massive strings, cannot be 
distinguished  by tensions at rest that point at their origin.

Second, components of the curvature tensor of MCS spacetime near the string worldsheet behave as a distribution. Curvature
singularities associated to the parabolic holonomies are analogs of conical singularities.

Third,  massless strings generate perturbations of the velocities of bodies resulting  in overdensities of matter. 
The strings also shift energies of photons, and may yield additional 
anisotropy of cosmic microwave background, if we consider MCS in a cosmological context. These effects of MCS are direct
analogs  of, respectively,  wake effects \cite{Brandenberger:2013tr} and the Kaiser-Stebbins effect \cite{Stebbins:1987va}, \cite{Sazhina:2008xs} 
known for common cosmic strings.  High energy MCS, for which $EG/c^4 \simeq 1$, 
where $G$ is the Newton coupling and $c$ is a speed of light, can be discriminated from massive strings.

Fourth, the worldsheet  of a MCS  is a geodesic null hypersurface, which belongs to an event horizon of the string. All events
which happen above the horizon cannot affect the string, while all events below the horizon are casually independent of the string. 
Mutual transformations of trajectories passing the string from different sides can be set on the string horizon 
\cite{Fursaev:2017aap}.
The presence of the horizon results in a sharp shape of spots of the CMB anisotropy caused by MCS.

The aim of the present work is to describe massless cosmic strings in curved backgrounds and extend  the method 
developed in \cite{Fursaev:2017aap}. Our main interest is in MCS  in expanding Universe and their observational effects.

The key problem is how to take into account the backreaction caused by the strings and ensure the required 
holonomy on the string worldsheet. For massive or massless cosmic strings
in a flat spacetime this problem is resolved with the help of isometries. For example, one can use the fact that the metric is axially symmetric
and change periodicity of the polar coordinate around a massive string to create an angle deficit related to the string tension \cite{Vilenkin:2000jqa}.
In case of massless strings parabolic isometries allow one to set a global  transformation of trajectories on the string 
horizon \cite{Fursaev:2017aap}. 

In this work, we construct massless cosmic strings moving in the de Sitter universe. Cosmic strings in the de Sitter 
spacetime have immediate cosmological implications. The de Sitter geometry fits the inflationary stage and it may serve
as an approximation for the present Universe. The de Sitter spacetime is maximally symmetric. One of its 
isometries  is an analog of the parabolic group, which we use to set coordinate transformations on the string horizon and ensure the required holonomy around the string.
This means that in the given model the backreaction effects of MCS are global and  metric around the string is locally de Sitter.

The paper is organized as follows. In Section 2 we start by summarizing construction  \cite{Fursaev:2017aap} of  background geometry 
of massless strings in a flat spacetime.  By this we mean establishing a set of rules for trajectories of particles and rays,  and, more generally,
for tangent bundles on this background. The rules are formulated at the string horizon.
Then we present the construction of MCS-de Sitter spacetime. We first do it
in terms of embedding of the de Sitter geometry in a 5D Minkowsky spacetime $R^{1,4}$.
MCS in the de Sitter spacetime are considered as an intersection of the de Sitter geometry and a flat massless 2D brane in $R^{1,4}$, an analog
of a straight MCS in $R^{1,3}$. By setting transition rules on the horizon of the brane along the lines of \cite{Fursaev:2017aap} one ensures the required holonomy.  To make a link to cosmological applications circular MCS are studied in a flat de Sitter universe. The strings are stretched along circles of the de Sitter (Hubble) radius. In Sec. 3 we discuss an image of the string seen by a freely moving observer in the de Sitter universe, a hypothetical case when the string emits light. We show that, like in case 
of straight MCS in Minkowsky space, a de Sitter observer sees the string as a moving circle of increasing radius.
Physical effects generated by MCS in the de Sitter spacetime are analyzed in Section 4. We focus on creation of regions of overdensities
in the flat de Sitter universe and on shifts of the energy of photons. The overdensities occupy wedge-shaped regions 
with a hole inside. Photons with shifted energies make spots which fill in the string image.  Summary and conclusions are given  in Sec. 5.

\section{Constructing MCS spacetimes}\label{s2}
\subsection{MCS-Minkowsky spacetime}\label{mink}
\setcounter{equation}0

We start with the definition of the parabolic subgropup of the Lorentz transformations. The holonomy around a MCS
is its element.  If $t,x,y,z$ are coordinates in Minkowsky spacetime $R^{1,3}$, the parabolic transformations (also known as null rotations),
$(x')^\mu=M^\mu_{~\nu}\left(\lambda\right)x^\nu$,  look as:
\begin{equation}\label{1.1}
u'=u~~,~~
v'=v+2\lambda y+\lambda^2u~~,~~
y'=y+\lambda u~~,~~z'=z~~,
\end{equation}  
where $u=t-x$, $v=t+x$ are light-cone coordinates and $\lambda$ is some real parameter. One can check that (\ref{1.1}) make a 
one-parameter group, $M(\lambda_1)M(\lambda_2)=M(\lambda_1+\lambda_2)$. 

Consider a massless string  stretched along the $z$-axis,
which moves along the $x$-axis at $y=0$. A parallel transport of a vector $V$ along a closed contour around the string 
should result in a non-trivial rotation, $V'=M(\lambda)V$, where $\lambda=8\pi G E$ and $E$ is an energy of the string  
per unit length \cite{vandeMeent:2012gb}.
The string worldsheet is $u=y=0$. It is a fixed point set of the parabolic  transformations.  

The spacetime around a MCS is locally Minkowsky. Because of the holonomy
Minkowsky-like coordinates  cannot be introduced globally on MCS geometry.  A straightforward introduction of such coordinates
results in delta-function-like singularities in the metric at $u=0$ \cite{Barrabes:2002hn}. 
The hupersurface $u=0$ is the event horizon of the string.

In practise,  the definition of MCS spacetime is equivalent to a set of rules to describe trajectories of particle or light rays near the string.
These rules can be also extended to fields (fibre bundles). The method suggested in \cite{Fursaev:2017aap} is the following.
The MCS spacetime is decomposed onto two parts: below, $u<0$, and above, $u>0$, the string horizon $u=0$.  Let us call trajectories at $u<0$ and $u>0$ 
ingoing and outgoing trajectories, respectively. Since ingoing trajectories  are casually independent of the string, they behave as in Minkowsky spacetime.

To describe outgoing trajectories, one introduces two types of coordinate charts: $R$- and $L$-charts, with cuts on the horizon either on the left
or on the right to the string, respectively. The string horizon is considered as a Cauchy hypersurface where initial data for 
outgoing trajectories are determined. These initial data are related to ingoing trajectories. Descriptions based on $R$- or $L$-charts are equivalent.
The choice of the chart is a matter of convenience depending on a trajectory of an observer.

For $R$-charts the position of the cut is $u=0, y<0$. 
The initial data  for an outgoing trajectory (coordinates and 4-velocities)  are
\begin{equation}\label{1.2}
x^\mu=M^\mu_{~\nu}\left(\lambda\right)\bar{x}^\nu\mid_{u=0,y<0}~~,
~~u^\mu=M^\mu_{~\nu}\left(\lambda\right)\bar{u}^\nu\mid_{u=0,y<0}~~.
\end{equation}  
\begin{equation}\label{1.3}
x^\mu=\bar{x}^\mu\mid_{u=0,y>0}~~,
~~u^\mu=\bar{u}^\mu\mid_{u=0,y>0}~~,
\end{equation}  
where $\bar{x}^\mu$, $\bar{u}^\mu$ are the coordinates and velocities of the corresponding ingoing trajectory when it reaches the horizon,
and $\lambda=8\pi GE$.   On $R$-charts the 'right' trajectories ($y>0$) behave smoothly across the horizon and, in particular,
the 'right' geodesics are just straight lines. The `left' trajectories  ($y<0$) are transformed after the horizon with respect to the `right' ones, although
mutual orientation of `left' trajectories does not change .  Let us emphasize that definition of `left' and `right' trajectories is invariant with respect to 
the parabolic transformations, since $y$ coordinate does not change under (\ref{1.1}) at the horizon.

Once the above rules are defined for trajectories, they can be extended to tangent vector bundles over the string spacetime. 
That is, we require that components of all vector fields satisfy the same transition rules (\ref{1.2}), (\ref{1.3})  across the horizon.
Since any closed contour around the string is made of `left' and `right' trajectories, it is clear from (\ref{1.2}), (\ref{1.3}) that a parallel transport  
of a vector along this contour is equivalent to a parabolic rotation with $M(\lambda)$. This guarantees the correct holonomy.

Finally, to make sure that discontinuities on the cut are coordinate discontinuities, and they are not physical, we require that
similar rules are introduced for all tensor structures and for fibre bundles (fields) over the string spacetime. 

The $L$-charts are dual to $R$-charts. They are smooth everywhere except the 
right cut on the horizon, $u=0, y>0$, where 'right' outgoing trajectories experience transformation
by the inverse matrix,
\begin{equation}\label{1.4}
x^\mu=\bar{x}^\mu\mid_{u=0,y<0}~~,
~~u^\mu=\bar{u}^\mu\mid_{u=0,y<0}~~,
\end{equation}  
\begin{equation}\label{1.5}
x^\mu=M^\mu_{~\nu}\left(-\lambda\right)\bar{x}^\nu\mid_{u=0,y>0}~~,
~~u^\mu=M^\mu_{~\nu}\left(-\lambda\right)\bar{u}^\nu\mid_{u=0,y>0}~~.
\end{equation}  
Description in terms of $R$ and $L$ charts are equivalent, since they are related by a global parabolic transformation
of the MCS spacetime at $u>0$.
 
Since the non-trivial holonomy is present, the Riemann curvature tensor must behave as a distribution at $u=y=0$.
This is indeed the case: as can be checked  \cite{Fursaev:2017aap}, the Riemann tensor has a single non-vanishing component
$R_{uyuy}=\lambda \delta(u)\delta(y)$. Analogous property of the Riemann tensor is known in case of conical singularities \cite{Fursaev:1995ef}.    The stress-energy tensor of the massless string,  which can be obtained 
in the ultrarelativistic limit of a massive string, is $T^{\mu\nu}=E\delta(u)\delta(y)u_s^\mu u_s^\nu$,
where $u_s^\mu$ is 4-velocity of the string, $u_s^2=0$. One can check that $T^{\mu\nu}$ stands in the right hand side of the Einstein equations
which take into account distributional property of MCS spacetime. Therefore, one has the explicit solution of the backreaction problem.

For further applications it is convenient to rewrite parabolic transformations (\ref{1.1}) in a form independent on the choice of coordinates. If $m$ is a unit spacelike vector orthogonal to the string and its velocity, transformation (\ref{1.1}) of a vector $V$ at the string worldsheet
can be written as follows \cite{Fursaev:1995ef}:
\begin{equation}\label{1.6}
M(\lambda)V=V-{\lambda^2 \over 2}u_s (u_s \cdot V)-\lambda m (u_s \cdot V) +\lambda u_s (m \cdot V)~~.
\end{equation}
We use notation $(a \cdot b)=a_\mu b^\mu$.

Definition of string velocity $u_s$ is related to a given frame of reference. If one boosts the frame along the string direction of motion, 
$u_s$ and $\lambda$ are changed to $e^\beta u_s$, $e^{-\beta}\lambda$, where $\beta$ is a parameter connected with the coordinate velocity.

\subsection{MCS-de Sitter spacetime in terms of embedding}\label{mcsds}

Our aim now is to extend construction of MCS spacetime to strings moving in an expanding universe. We require that:
i) the worldsheet
of a massless string is a null geodesic surface in a curved geometry, that is each point on the string moves along a null geodesic, ii) a parallel transport of
a vector along a small contour around a point on the string worldsheet results in the same transformation by a parabolic group
as in the flat spacetime.

In this paper we consider MCS in de Sitter spacetime, where such null surfaces can be easily and explicitly constructed.
Note that solutions for massive strings in de Sitter spacetime (whose worldsheets are extremal surfaces) are known for a long time,  see e.g. \cite{Linet:1986sr}-\cite{deVega:1994yz}.

The advantage of the de Sitter geometry is in its isometries.  This geometry can be embedded in a five-dimensional Minkowsky spacetime $R^{1,4}$ (with coordinates $X^K$),
\begin{equation}\label{2.1}
-X_0^2+X_1^2+X_2^2+X_3^2+X_4^2=-UV+Y^2+X_2^2+X_3^2=H^{-2}~~.
\end{equation}
Here $1/H$ is the de Sitter radius, $U=X^0-X^1$,  $V=X^0+X^1$, $Y=X_4$. Consider in $R^{1,4}$ a null hypersurface $U=Y=0$, 
which can be interpreted as worldsheet of a massless brane. Its intersection 
with (\ref{2.1}) is a null surface in (\ref{2.1}). It is a space product $R^1\times S^1$ whose cross sections are circles $X_2^2+X_3^2=H^{-2}$. Points on the circles move along null geodesics on the de Sitter spacetime. We interpret this null surface as a worldsheet of a massless circular cosmic string
moving in de Sitter spacetime.

The event horizon of the massless brane is $U=0$. Its intersection with the de Sitter spacetime is the event horizon of the MCS .
The string horizon is $R^1\times S^2$ whose sections are 2-spheres. MCS are described more explicitly in Sec. \ref{flatU} in the model of a
flat de Sitter universe.

We now need to construct a spacetime around the MCS.  To ensure the required holonomy on the string worldsheet we assume 
that a parabolic holonomy, say $M_5(\lambda)$, exists on
the brane $U=Y=0$ in $R^{1,4}$.  The coordinate transformations $X'=M_5(\lambda)X$, which leave  the surface $U=Y=0$ invariant, 
are defined as in  (\ref{1.1}),
\begin{equation}\label{2.2}
U'=U~~,~~
V'=V+2\lambda Y+\lambda^2 U~~,~~
Y'=Y+\lambda U~~,~~X_2'=X_2~~,~~X_3'=X_3~~.
\end{equation} 
These transformations  make a parabolic subgroup of $O(1,4)$. When restricted to (\ref{2.1}) they become isometries of the de Sitter spacetime.
They map a vector $V$ in a tangent space at a point $X$ to a tangent vector $V'=M_5(\lambda)V$ at a point $X'=M_5(\lambda)X$.
(Note that a vector with components $X^K$ is orthogonal to the tangent space, and (\ref{2.2}) holds scalar products in $R^{1,4}$ .) 

Since $M_5(\lambda)$ does not move $X^K$ at the string worldsheet, it acts there precisely as 
in (\ref{1.6}). To see this one can write 
\begin{equation}\label{2.3}
M_5(\lambda)V=V-{\lambda^2 \over 2}u_b (u_b \cdot V)-\lambda m_b (u_b \cdot V) +\lambda u_b (m_b \cdot V)~~
\end{equation}
at $U=Y=0$. Here $u_b$ is a null 4-velocity of the brane and $m_b$ is a spacelike unit vector orthogonal to the brane and its direction of motion,
 $(A \cdot B)\equiv A^KB_K$.
If one considers (\ref{2.3}) for a tangent vector $V$ at the string world sheet, $u_b$ in  (\ref{2.3})  can be replaced to $u_s$
(since the brane and the string on the de Sitter spacetime move along the same trajectory), and $m_b$ to $m$ (since 
$m_b$ is tangent, $(m_b\cdot X)=Y=0$).

The MCS-de Sitter universe is then constructed by setting transition rules for trajectories in $R$- or $L$- coordinate charts.
The $R$ chart is smooth everywhere except for a left cut along the string horizon $U=0,Y<0$.
The transition rules on the horizon for a tangent vector field $V$ can be defined in analogy  with  (\ref{1.2}),(\ref{1.3}). 
Let ${\bar V}$ be the vector field when the horizon is approached from below. The transformed field is
\begin{equation}\label{2.4}
V(X)=M_5(\lambda){\bar V}({\bar X})\mid_{U=0,Y<0}~~,~~
X=M_5(\lambda){\bar X}
\end{equation}  
\begin{equation}\label{2.5}
V(X)={\bar V}(X)\mid_{U=0,Y>0}~~.
\end{equation}  
In $L$ charts, where the cut is $U=0,Y>0$ the rules are
\begin{equation}\label{2.6}
V(X)={\bar V}(X)\mid_{U=0,Y<0}~~,
\end{equation}   
\begin{equation}\label{2.7}
V(X)=M_5(-\lambda){\bar V}({\bar X})\mid_{U=0,Y>0}~~,~~
X=M_5(-\lambda){\bar X}~~.
\end{equation}
The rules above transform tangent space to a tangent space.

The Riemann curvature tensor of the constructed MCS spacetime is  the sum,
\begin{equation}\label{2.8}
R_{abcd}=R^{\mbox{\tiny DS}}_{abcd}+\tilde{R}_{abcd}~~,
\end{equation}
of regular, $R^{\mbox{\tiny DS}}$, and singular,  $\tilde{R}$, parts.
The regular part is the Riemnn tensor of the de Sitter geometry.  The singular part  appears as a result 
of the holonomy around the string, it behaves as a distribution at $U=Y=0$. To compute $\tilde{R}$ 
it is enough to consider a domain very close to the string worldsheet and use results of Sec. \ref{mink}. 
It is convenient to choose  a tetrade basis which consists of $u_s,m$ and two additional tangent vectors $v,l$, such that  $v$ is null and $(v\cdot u_s)=-2$,
$l$ is spacelike, unit and tangent to the string ($(l\cdot m)=0$ and $l,m$ are orthogonal to $u,v$). The tetrade is defined at the worldsheet.
Results of MCS in flat spacetime and the fact that $M_5(\lambda)$ acts at the worldsheet  as $M(\lambda)$ indicate that
the singe non-vanishing component of $\tilde{R}$ in this basis is $\tilde{R}_{umum}=\lambda \delta_{WS}(x)$.
The delta-function $\delta_{WS}(x)$ is properly normalized and has the singularity on the worldsheet.
The stress-energy energy tensor of the string follows from the right hand side of the Einstein equations.
The only non-vanishing component is $T_{uu}=E \delta_{WS}(x)$, 
where the parameter $E=\lambda/(8\pi G)$ can be interpreted as the energy of the string per unit length.

\begin{figure}[h]
\begin{center}
\includegraphics[height=6.5cm,width=9cm]{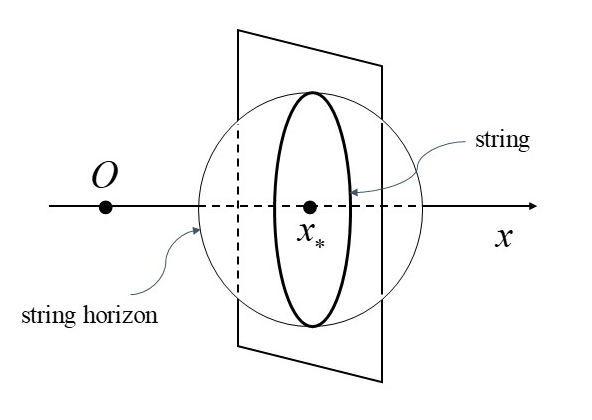}
\caption{\small{a massless cosmic string described in Sec. \ref{flatU} is shown. The center 
of the string horizon sphere is at $x_\star$, the position of the observer at the center
of coordinates is at the point $O$.}}
\label{f2} 
\end{center}
\end{figure}

\subsection{Circular MCS in flat de Sitter universe}\label{flatU}

In what follows we consider strings in a flat de Sitter universe with the metric
\begin{equation}\label{3.1}
ds^2=-dt^2+a^2(t)(dx^2+dy^2+dz^2)~~,
\end{equation}
where $a(t)=e^{tH}$.  Relations to coordinates $X^K$ which lead to (\ref{3.1})  are
\begin{equation}\label{3.3}
X^0=H^{-1}\sinh \chi+\frac 12 Hr^2e^{\chi}~~,~~X^1=x~ e^{\chi}~~,~~X^3=z~ e^{\chi}~~,~~\chi\equiv tH~~,
\end{equation}
\begin{equation}\label{3.4}
X^2=\sin \alpha \left(H^{-1}\cosh \chi-\frac 12 Hr^2e^{\chi}\right)+\cos \alpha ~y~ e^{\chi}~~,
\end{equation}
\begin{equation}\label{3.5}
Y=\cos \alpha \left(H^{-1}\cosh \chi-\frac 12 Hr^2e^{\chi}\right)-\sin \alpha~ y~ e^{\chi}~~,
\end{equation}
$0\geq \alpha <\pi/2$.
In the given parametrization  trajectory of an observer at the center of coordinates (\ref{3.1}), $x=y=z=0$, is given by
\begin{equation}\label{3.2}
X^0(t)=H^{-1}\sinh \chi~~,~~X^a(t)=M^a H^{-1}\cosh \chi~~,~~a=1,2,3,4~~,
\end{equation}
where $M$ is a unit vector with 2 nonvanishing components $M^Y=\cos \alpha$, $M^2=\sin \alpha$.

The above coordinates cover the domain $X^0+\sin\alpha~ X^2+\cos\alpha~ Y >0$ of the full de Sitter geometry, 
they are restricted by the past event horizon of an observer at the center of coordinates. In the model we consider the string moves inside 
the given domain. 
To see this we note parametrization of the $U$-coordinate 
\begin{equation}\label{3.6}
U={He^\chi \over 2}\left( (\vec{x}-\vec{x}_\star)^2-H^{-2}e^{-2\chi}\right)~~,
\end{equation}
where $\vec{x}_\star$ has components $x_\star=1/H,y_\star=z_\star=0$.
Equation for the string horizon $U=0$ is 
\begin{equation}\label{3.10}
(\vec{x}-\vec{x}_\star)^2=H^{-2}e^{-2\chi}
\end{equation}
This is a sphere which in the physical (expanding) coordinates $e^{\chi}x^i$ has 
a constant Hubble radius $H^{-1}$. That is, the string horizon is the cosmological horizon 
for an observer at a point $\vec{x}_\star$.

To visualize the string worldsheet (\ref{3.5}) can be written as
\begin{equation}\label{3.8}
Y=- \cos\alpha ~U-e^\chi \left(\vec{p}\cdot (\vec{x}-\vec{x}_\star)\right)~~,
\end{equation}
where $(\vec{a}\cdot \vec{b})=a_ib_i$ is the 3D scalar product and $\vec{p}$ is a unit vector with 
components $p_x=\cos\alpha, p_y=\sin\alpha,p_z=0$. Therefore, the string worldsheet $U=Y=0$ is the intersection 
of sphere (\ref{3.10}) and the plane 
\begin{equation}\label{3.9}
\left(\vec{p}\cdot (\vec{x}-\vec{x}_\star)\right)=0~~.
\end{equation}
The plane with normal vector $\vec{p}$ goes through the center of the sphere. The string is a large circle of the cosmological horizon,
see Fig. \ref{f2}.
We call (\ref{3.9}) the plane of the string. 

In the expanding universe the plane moves with respect to observers with fixed coordinates $x^i$, except for observers located on the plane itself.
For example,  the string plane moves with respect to an observer at the center of coordinates
along $x$ axis and it is tilted to the axis with the angle $\alpha$ defined in (\ref{3.4}), (\ref{3.5}).

The above observer, being inside the string horizon in the past, crosses the horizon at the moment $t=0$.
It is easy to see form (\ref{3.8}) that he is a `right' observer ($Y>0$), if $\cos\alpha >0$, and  he is `left', if $\cos\alpha <0$.
The future event horizon of the observer is $X^0-\sin\alpha~ X^2-\cos\alpha~Y =0$, which is the sphere $|\vec{x}|=e^{-\chi}/H$.
If $\alpha\neq 0$ a segment of the string stays inside the observer's horizon until the moment $t=H^{-1}\ln (2\sin\alpha)$

\section{Image of MCS}\label{image}
\subsection{Images of strings and branes in flat spacetimes}
\setcounter{equation}0

If a string emitted light, how its image would
look like for an observer? Let $\vec{n}$ be a unit vector at the observer's location tangent to his past light cone. By the definition,
the image of the string is a set of directions, determined by $\vec{n}$, which belongs to that part of the light cone 
which intersects the string worldsheet.

We describe the image of a MCS for an observer at the center of coordinates (\ref{3.1}). We use embedding 
of the given problem in $R^{1,4}$ and results \cite{Fursaev:2017aap}.  Consider a 
straight MCS in flat spacetime described in Sec. \ref{mink}.
Equation for the past light cone of an observer with coordinates ${\vec x}_o$ at a moment $t_o$ is 
\begin{equation}\label{4.1}
{\vec x}(t)={\vec x}_o+(t-t_o){\vec n}~~,~~t<t_o~~.
\end{equation}
For rays which make an image of the string: $y(t_e)=0$, $x(t_e)=t_e$, where $t_e$ is a moment of emission from the string.
If the position of the observer in the $(x,y)$ plane is $x_o=0$, $y_o=b$, where $b>0$,
equation for $\vec{n}$  is derived in the following form \cite{Fursaev:2017aap}:
\begin{equation}\label{4.2a}
\sin\theta\cos(\varphi-\hat{\theta})=\cos \hat{\theta}~~.
\end{equation}
Here $\theta$, $\varphi$ are  common angles on the sphere  $\vec{n}^2=1$, and
\begin{equation}\label{4.3}
\cos \hat{\theta}={b \over \sqrt{b^2+t_o^2}}~~,
\end{equation}
$0<\hat{\theta}<\pi/2$. Note that position of the observer along the string does not matter.

\begin{figure}[h]
\begin{center}
\includegraphics[height=5cm,width=10cm]{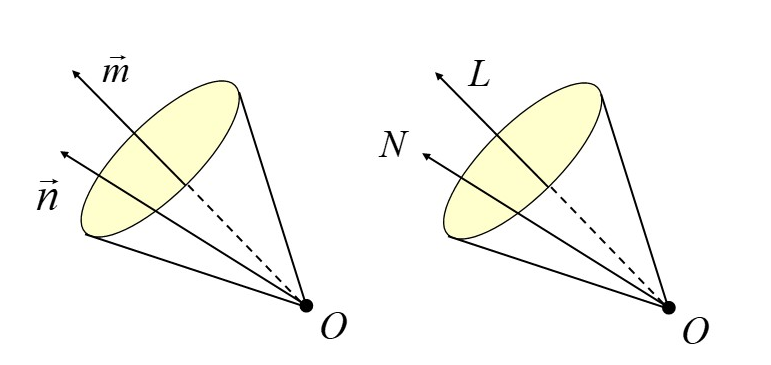}
\caption{\small{the visibility cones of the string for an observer at the point $O$ in four (the left figure) and five (the right figure) 
spacetime dimensions.}}
\label{f1} 
\end{center}
\end{figure}

For further purposes we rewrite (\ref{4.2a}) in a geometrical form,
\begin{equation}\label{4.2}
({\vec n} \cdot {\vec m})=\cos \hat{\theta}~~,
\end{equation}
where ${\vec m}$ is a unit vector with components $m_x=\cos \hat{\theta}, m_y=\sin \hat{\theta},m_z=0$.   
Directions of $\vec n$ determined by (\ref{4.2}) make a cone with an axis along $\vec m$, see Fig. \ref{f1}. We call it the visibility cone of the string. 
Its axis $\vec m$ is orthogonal to the string.
An image of the straight MCS on a sky for the given observer is determined by the visibility cone, and it is a circle.  
There is no contradiction between shape of the image
and shape of the string. The image is made of rays emitted by different points of the string at different moments.

When the observer crosses the string horizon, $\vec m$ is directed along the coordinate velocity of the string.
At large times, $t_o\gg b$, the cone converts to a plane,  and $\vec m$ becomes orthogonal to  the velocity of the string.  

Equation (\ref{4.2}) can be generalized straightforwardly to higher dimensions, for example, to describe an image of a flat massless brane in $R^{1,4}$ considered in Sec.\ref{mcsds}. 
The past light cone of an observer with coordinates $X^L_o$ in $R^{1,4}$  is
\begin{equation}\label{4.6}
X^a(T)=X^a_o(T_o)+(T-T_o)N^a~~,~~X^0(T)=T~~.
\end{equation}
$T_o$ is the time of the observation, $N^a$ is a unit vector ($a=1,2,3,Y$), analogous to ${\vec n}$, which sets directions of 
photons emitted from the brane.
The visibility cone of the brane in $R^4$ is determined as, see Fig. \ref{f1},
\begin{equation}\label{4.4}
(N \cdot L)=\cos \Theta~~.
\end{equation}
Here $L$ is a unit vector, analogous to  ${\vec m}$, with components
$L^0=0,L^1=\cos\Theta, L^Y=\sin\Theta,L^2=L^3=0$, see notation for the indexes in (\ref{2.1}). 
Cone's axis $L$ is orthogonal to the brane.
Suppose location of the observer in $X^1,Y$ plane is $X^1_o=0,Y_o\neq 0$, then
\begin{equation}\label{4.5}
\cos\Theta={Y_o \over \sqrt{Y_o^2+T_o^2}}={\cos\alpha\cosh \chi \over \sqrt{\cos^2\alpha\cosh^2 \chi+\sinh^2 \chi}}~~,
\end{equation}
 see (\ref{4.3}). The trajectory of the considered de Sitter observer is given by (\ref{3.2}). Therefore,  in (\ref{4.5})
we put $T_o=H^{-1}\sinh \chi$, $Y_o=H^{-1}\cos\alpha\cosh \chi$.
 
\subsection{String image for a de Sitter observer}\label{im}

Equation (\ref{4.4}) is the visibility cone of a flat brane as seen by 5-dimensional observer in $R^{1,4}$. We need
the picture for de Sitter observers. We consider the observer at the center of coordinates (\ref{3.1}). 
Trajectories of light rays (\ref{4.6})  which are tangent to the de Sitter spacetime satisfy additional restrictions:
one can show that (\ref{4.6}) is tangent, if
\begin{equation}\label{4.7}
(M\cdot N)=\tanh \chi~~,
\end{equation}
where $M$ is unit vector for observer's trajectory (\ref{3.2}). 

Let   $U_{o}$ and $U_{\gamma}$ be five-velocities of the observer and the light ray, respectively ,
\begin{equation}\label{4.8}
U^0_{o}=\cosh \chi~~,~~U^a_{o}=\sinh \chi M^a~~,~~U^0_{\gamma}=1~~,~~U^a_{\gamma}=N^a~~~.
\end{equation}
By using (\ref{4.7}), (\ref{4.8}) one finds that at the point of emission $(U_{o}\cdot U_{\gamma})=-1/\cosh\chi$.
One can decompose $U_{\gamma}$ there as 
\begin{equation}\label{4.9}
U_{\gamma}={1 \over \cosh\chi} (U_{o}+n)~~~,~~~(U_{o}\cdot n)=0~~,~~(n\cdot n)=1~~.
\end{equation}
Since $U_{o}$, $U_{\gamma}$ in (\ref{4.9}) lie in the tangent plane, so does vector $n$. 
In fact, (\ref{4.9}) is a decomposition of $U_{\gamma}$ in the frame of reference of the de Sitter observer,
where $n$ determines the observer's past light cone. One can check that 
\begin{equation}\label{4.10}
n^0=0~~,~~n^a=\cosh \chi N^a-\sinh \chi M^a~~.
\end{equation}
One can also get $n$ by projecting $N$ to a part of the tangent space which is orthogonal to velocity of the observer:
\begin{equation}\label{4.11}
P^K_L N^L={n^K \over \cosh\chi}~~,~~P^K_L=\delta^K_L-H^2X_o^KX_{oL}+U_o^KU_{oL}~~.
\end{equation}
Here we used (\ref{4.7}), (\ref{4.8}), (\ref{4.10}). Analogously, one can project the axis vector $L$ of 5-dimensional cone (\ref{4.4}):
\begin{equation}\label{4.12}
P^K_Q L^Q=L^K-\cos\alpha \sin\Theta M^K~~,
\end{equation}
where we used $(L\cdot M)=L^YM^Y=\cos\alpha \sin\Theta$. The axis of the visibility cone of the de Sitter observer is the normalized 
projection (\ref{4.12}),
\begin{equation}\label{4.13}
m={L-\cos\alpha \sin\Theta M \over \sqrt{1-\cos^2\alpha \sin^2\Theta}}~~.
\end{equation}
By using  (\ref{4.5}), (\ref{4.10}),(\ref{4.13}) we finally find the equation for the visibility cone of the MCS in the de Sitter universe:
\begin{equation}\label{4.14}
(n\cdot m)=\cos \hat{\theta}~~,~~\cos \hat{\theta}={\cos\alpha \over \sqrt{\cos^2\alpha \cosh^2\chi+\sin^2\alpha \sinh^2\chi}}~~.
\end{equation}
MCS in the de Sitter universe look as in a flat spacetime. They are circles which grow with time. 
When the observer crosses
the string horizon at $t=0$, $n$ is directed along $m$, and the MCS looks as a point.  When $t\ll 1/H$ the angle 
$\hat{\theta}$ increases as $tH/\cos\alpha$, where $t$ is the proper time of the observer.
At large times $t\geq 1/H$ vectors $n$ and $m$ become orthogonal exponentially fast, the cone converts to the plane. 
$n$ and $m$ are always orthogonal when the observer is in the plane of the string, $\alpha=\pi/2$.

It is interesting to discuss rotation of $m$ with time. Since $m$ is a tangent vector one can compute its components 
along $x,y,z$ directions in coordinates  (\ref{3.1}). Three unit vectors $e^L_i=e^{-\chi}X^L_{,i}$ which determine these directions at the center of coordinates
are $e_x^L=\delta_1^L$, $e_y^L=\cos\alpha \delta_2^L-\sin\alpha \delta_Y^L$, $e_z^L=\delta_3^L$. In this basis
$m$ has the following non-vanishing components $m_i=(m\cdot e_i)$:
\begin{equation}\label{4.15}
m_x={\cos\alpha \cosh\chi \over \sqrt{\cos^2\alpha \cosh^2\chi+\sin^2\alpha \sinh^2\chi}}~~,~~
m_y=-{\sin\alpha \sinh\chi \over \sqrt{\cos^2\alpha \cosh^2\chi+\sin^2\alpha \sinh^2\chi}}~~.
\end{equation}
When the observer crosses the string horizon he sees the string as a point in the direction $m=(1,0,0)$. At large times components
are $m_x=\cos\alpha, m_y=-\sin\alpha$.

We described the string image for a particular observer  at the center of coordinates. Analogous analysis can be done for other observers.
It is clear that images depend on observer's location. An observer at the center of the string horizon sphere never sees the string (observer's
cosmological horizon
coincides with the string horizon).

\section{Physical effects}\label{eff}
\subsection{Parabolic holonomy and a Killing vector}
\setcounter{equation}0

Consider physical effects caused by  MCS in flat de Siter universe (\ref{3.1}).
We describe these effects from the point of view of `right' observers whose trajectories when crossing the string horizon satisfy 
conditions $Y>0$ or $\left(\vec{p}\cdot (\vec{x}-\vec{x}_h)\right)<0$, see  (\ref{3.8}).
The effects we are interested in are transformations of trajectories
crossing horizon from the left, $Y<0$.

We use a $R$ coordinate chart introduced in Sec. \ref{mcsds}. Transformation of `left' trajectories  in this chart are
given by (\ref{2.4}), (\ref{2.5}).  Since we work in coordinates
$x^\mu=\{t,x^i\}$  we should know how transformations of $X^K$, see (\ref{2.2}),  generate transformations of $x^\mu$.
The definition  is
\begin{equation}\label{5.1}
M^L_K(\lambda) X^K(x^\mu)=X^L(y^\mu)~~.
\end{equation}
The above relation between $y^\mu$ and $x^\mu$ is essentially non-linear. We demonstrate how calculations can be done at small $\lambda$ (low energy MCS) in the linear approximation. By using (\ref{2.2}) we define the Killing vector field $\zeta^L$:
\begin{equation}\label{5.2}
\delta X^L=X^L-\bar{X}^L\equiv \lambda \zeta^L +O(\lambda^2)~~,
\end{equation}
where $X=M_5(\lambda) \bar{X}$, and  the tangent Killing vector  field $\zeta^\mu$, which generates
transformations of $x^\mu$,
\begin{equation}\label{5.3}
\delta x^\mu=x^\mu-\bar{x}^\mu \equiv \lambda \zeta^\mu +O(\lambda^2)~~.
\end{equation}

\begin{figure}[h]
\begin{center}
\includegraphics[height=8cm,width=12cm]{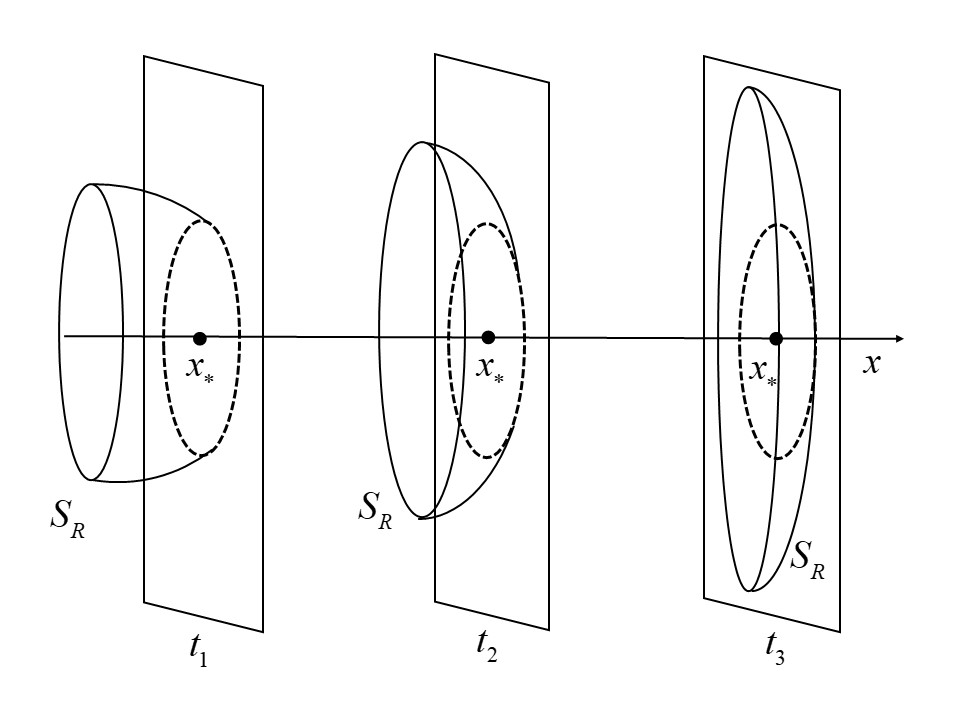}
\caption{\small{the evolution of the region of overdensities in expanding coordinates at moments
of time $t_1<t_2<t_3$. The string moves from the left to the right, and it is shown as a dashed ellipse. 
The overdensities  are located inside 
a region between parabolic surface $S_R$ and a part of the sting plane outside the string.}}
\label{f3} 
\end{center}
\end{figure}
$\zeta^L$ has only 2 non-vanishing components: $\zeta^V=2Y$, $\zeta^Y=U$.
By using (\ref{5.1}) on the tangent space one finds 
\begin{equation}\label{5.4}
\zeta^L=X^L_{~,\mu}\zeta^\mu~~,~~\zeta^\mu=g^{\mu\nu}X^L_{~,\nu}g_{LK}\zeta^K=g^{\mu\nu}(Y_{,\nu}U-YU_{,\nu})~~,
\end{equation}
\begin{equation}\label{5.8}
\zeta^0=-\left(\vec{p}\cdot (\vec{x}-\vec{x}_\star)\right)~~,~~\zeta^i=H(x^i-x^i_\star)\left(\vec{p}\cdot (\vec{x}-\vec{x}_\star)\right)-p^ie^{-\chi}U~~,
\end{equation}  
where $U$ is defined in (\ref{3.6}). One can see from (\ref{5.8}) that $\zeta^\mu$ satisfy the Killing equations, and 
$\zeta^\mu=0$ on the string worldsheet.

Now transition conditions (\ref{2.4}), (\ref{2.5}) in the $R$-chart for a vector field $V$ from a tangent vector bundle 
can be written as
\begin{equation}\label{5.6}
\delta V=-\lambda {\cal L}_\zeta V\mid_{U=0,Y<0}~~.
\end{equation}  
The change of the field on the left side of the string horizon is generated by the Lie derivative under the 
coordinate transformation $\delta x^\mu=\lambda \zeta^\mu$.
On the right side, $U=0,Y>0$, the field does not change, $\delta V=0$. Definition of the Lie derivative for vector fields, which we 
use, is
\begin{equation}\label{5.5}
{\cal L}_\zeta V^\mu=\zeta^\nu V^\mu_{~,\nu}-\zeta^{\mu}_{~,\nu}V^\nu~~.
\end{equation}
One should emphasize that (\ref{5.6}) holds when transformations are not large, $|\delta V/V|\ll 1$. Otherwise one should use
exact coordinate transformations (\ref{5.1}) and find corresponding changes for field components.

\subsection{Wake effects}\label{wake}

Consider a matter which is at rest with respect to coordinates $x^i$ of flat de Sitter universe (\ref{3.1}).
Such a matter moves freely along geodesics with 4-velocities $u^t=1,u^i=0$. The 4-velocities 
of left outgoing geodesics change to
\begin{equation}\label{5.7}
\delta u^i=-\lambda{\cal L}_\zeta u^i=-\lambda e^{-2\chi} p^i~~,~~\delta u^t=O(\lambda^2)~~,
\end{equation}
where $p^i$ is the unit normal vector of the string plane.  To get  (\ref{5.7}) we used (\ref{5.8}), (\ref{5.5}).  
In fact, (\ref{5.7}) can be considered as a transformation of a left trajectory in the entire region
behind the string horizon. It satisfies required conditions  (\ref{5.6}).

According to  (\ref{5.7}) the left matter after crossing the horizon starts to move with respect to the coordinate grid orthogonally to the string plane.
Its coordinate velocity $v^i=u^i/u^0$  is proportional to $8\pi G Ec$, where $E$ is the energy of the string, $c$ is the speed of light. 
This creates domains of overdensities when the `left' matter shifts to $R$-region.
The effect is completely analogous to the wake effect \cite{Brandenberger:2013tr} known for moving massive cosmic strings.  
Analogously, straight MCS in flat spacetime cause the left matter to move orthogonally to the string and its direction of motion \cite{Fursaev:2017aap}.

Overdensities for common strings and MCS in flat spacetime occupy a wedge-like region with the edge on the string \cite{Brandenberger:2013tr}.
A specific feature of MCS in expanding universe is that the shape of the region where overdensities are located is changing. This effect can be described with the help of  (\ref{5.7}).
Without loss of generality it is convenient to choose coordinates $x,y,z$, where the string plane is  $x=0$ ($p_x=1$). 
The center of coordinates can be placed at the point $\vec{x}_\star$, so that
string horizon sphere (\ref{3.10})  is given by equation $H^2\vec{x}^2=e^{-2\chi}$. The left trajectories are at $x>0$ and they move along the $x$ coordinate after crossing the horizon.

From the point of view of an $R$-observer the overdensities form in a region between two boundaries, $S_R$ and $S_L$. 
$S_L$ is a part of the string plane outside the string, $S_R$ is made of left trajectories
which start from the string worldsheet.
According to (\ref{5.7})  an outgoing left trajectory  of this kind is
\begin{equation}\label{5.8a}
x(t)={\lambda \over 2H}\left(e^{-2Ht}-e^{-2Ht_0}\right)~~.
\end{equation}
It crosses the string plane $x=0$ at the moment $t_0$ and moves to the $R$-region, $x<0$, with the velocity $v^x(t)=-\lambda e^{-2Ht}$.
One can relate $t_0$ with fixed coordinates $y$, $z$ of the given trajectory.  Since it crosses the horizon at $t=t_0$, one has
\begin{equation}\label{5.9}
H^2\rho^2=e^{-2Ht_0}~~,
\end{equation}
where $\rho^2=y^2+z^2$.  By taking into account (\ref{5.8a}), (\ref{5.9})
one can write equation for $S_R$ as
\begin{equation}\label{5.10}
x={\lambda \over 2H}\left(e^{-2Ht}-\rho^2H^2\right)~~.
\end{equation}
It is assumed that $H\rho\geq e^{-Ht}$. Thus, $S_R$ is a parabolic surface which ends on the string.

\begin{figure}[h]
\begin{center}
\includegraphics[height=6.5cm,width=6.5cm]{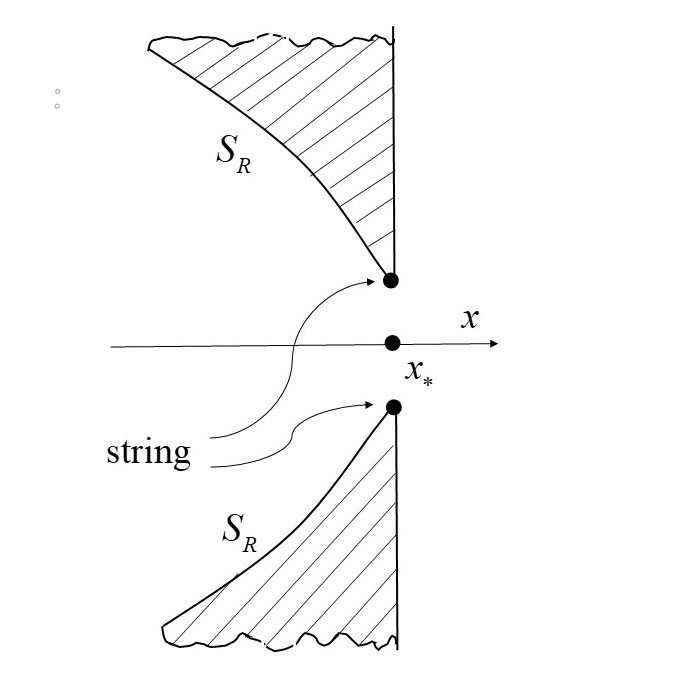}
\caption{\small{a cross-section of the region of overdensities (shaded area) by a plane going through the $x$ axis.  Near the string the region has  a wedge-like form such that the edge of the wedge lies on the string. String moves from the left to the right.}}
\label{f4} 
\end{center}
\end{figure}

The cosmic expansion causes the string to move
with respect to observers with fixed coordinates $\vec{x}$. To better understand
evolution of $S_R$ it is worth rewriting  (\ref{5.10}) in expanding coordinates, say 
$\hat{\vec{x}}(t)=e^{-tH}\vec{x}(t)$.
Equation (\ref{5.10}) becomes
\begin{equation}\label{5.11}
\hat{x}={\lambda(t) \over 2H}\left(1-\hat{\rho}^2H^2\right)~~.
\end{equation}
Here $\lambda (t)=4\pi G E(t)$, and $E(t)= e^{-Ht}E$ can be interpreted as a redshifted energy of the string (per unit length).  

Time evolution of the region of overdensities in expanding coordinates is shown on  Fig. \ref{f3}. 
The string is a circle of a constant radius which moves with respect to an observer outside the string plane.  Trajectories which cross the string horizon sphere on the right from the string plane acquire additional velocity
toward the string plane. The overdensities  appear when these trajectories cross the string plane.
Overdensities 
are  in a region with a disc-shaped hole inside.  Near the string it looks as a wedge
with the edge on the string, see  Fig. \ref{f4}. This form of overdensities is similar to what one has in case of moving massive strings
and MCS in a flat spacetime. The boundary of the wedge is made of parabolic surface $S_R$ and a part of the sting plane. At late times  the wedge is squeezed  toward the string plane.

\subsection{Shifts of energies of photons}

For $R$-observers, MCS in a flat spacetime change the energy of left photons leaving the string horizon \cite{Fursaev:2017aap}.
The same effect holds for MCS in the de Sitter universe.  Let $u_o$ be 4-velocity of an observer, and $k$ be a 4-momentum of a photon
registered by the observer. By following Sec. \ref{im} one can decompose the 4-momentum as (compare with  (\ref{4.9}))
\begin{equation}\label{5.12}
k=\omega(u_o+n)~~,
\end{equation}
where $\omega$ is a measured energy of the photon. 
Direction of motion of the photon in  (\ref{5.12}) is fixed by a unit vector $n$, which is orthogonal to 
$u_o$. 

Energy of the left photon as measured by an $R$-observer is shifted by the quantity
\begin{equation}\label{5.13}
\delta \omega =-(u_o\cdot \delta k)=\lambda {\cal L}_\zeta k^0~~,
\end{equation}
see (\ref{5.6}).
The energy of the photon in expanding universe  (\ref{3.1}) at the observation moment $t$ is $\omega(t)=\bar{\omega}/a(t)$, where $\bar{\omega}$ is a constant. This allows one to compute time derivatives. A straightforward computation then yields:
\begin{equation}\label{5.14}
\delta  k^0= -{\lambda k^0 \over a}\left(-\dot{a} \zeta^0+ \left(\vec{p}\cdot \vec{n}\right)\right)~~.
\end{equation}
We introduced  here a unit (in 3D Cartesian coordinates) vector $\vec{n}$ determined by 4-vector $n$, which sets the direction of the photon in 3D 
coordinates.
The relative shift of the energy can be written as a combination of two, effects,
\begin{equation}\label{5.15}
{\delta  \omega \over \omega} = {\delta  \omega_1 \over \omega}+ {\delta  \omega_2 \over \omega}~~,
\end{equation}
\begin{equation}\label{5.16}
{\delta  \omega_1 \over \omega} = -  \lambda(t)\left(\vec{p}\cdot \vec{n}\right)~~.
\end{equation}
\begin{equation}\label{5.17}
{\delta  \omega_2 \over \omega} = -  \lambda(t)~He^{Ht}\left(\vec{p}\cdot \vec{x}\right)
= \lambda(t)~z_\star\left(\vec{p}\cdot \vec{m}_\star\right)~~,
\end{equation}
where $\lambda(t) =8\pi G E(t)$, 
$E(t)$ is redshifted energy of the string, see Sec. \ref{wake}.
To get (\ref{5.16}), (\ref{5.17}) we used (\ref{5.8}), (\ref{5.14}). 
The observer is located at a point with coordinates $\vec{x}$. The center of coordinates  $\vec{x}=0$
is placed at the center of the string horizon sphere. A unit vector $\vec{m}_\star$ is 
at the location of the observer and is directed toward the center $\vec{x}=0$.
$z_\star$ is a redshift of the the center of the string horizon with respect to the observer.
It is easy to see that $He^{Ht}\vec{x}=-z_\star\vec{m}_\star$. Note that (\ref{5.16}), (\ref{5.17}) hold for $|\delta \omega|/\omega \ll 1$ since transformations (\ref{5.3}) are assumed to be small.

The shift $\delta  \omega_1/\omega$ has the same form as in case of MCS in a flat spacetime, see  \cite{Fursaev:2017aap}. 
For a straight MCS in a flat spacetime $\vec{p}$ in (\ref{5.16}) would be a unit vector orthogonal to the string worldsheet. For MCS in the de Sitter universe
$\vec{p}$ is directed orthogonally to the string plane.
The second term in (\ref{5.15}), $\delta  \omega_2/\omega$, is of a cosmological origin. It is an isotropic shift of energies of photons, 
independent of directions  
the photons come from. Note that, at late times $\omega_2$ dominates over $\omega_1$.

Equations  (\ref{5.16}), (\ref{5.17}) hold for left  photons detected by $R$-observers. Let 
us demonstrate that directions of left photons are inside the string image seen by observer, that is $n$ in
(\ref{5.16}) is inside the visibility cone of the MCS. We use results of Sec. \ref{im} which were obtained for an observer in the
equatorial plane of the string horizon sphere (i.e. with zero $y$ and $z$ coordinates). The string visibility cone is a projection
of the visibility cone of a 5-dimensional observer.  Let $n_s$ and $N_s$ be unit vectors which define directions of a photon
emitted by the string in 4- and 5-dimensional spacetimes, as it is was defined in Sec. \ref{im}.  Let $m$ and $L$ be unit vectors
which define the axis of the corresponding cones.  Equations for the cones are $(n_s \cdot m)=\cos \hat{\theta}$, see (\ref{4.14}),
and $(N_s \cdot L)=\cos \Theta$, see (\ref{4.4}), and Fig. \ref{f1}. We need to prove that
\begin{equation}\label{5.18}
(n\cdot m)\leq (n_s\cdot m)~~.
\end{equation}
To this aim we note that $n=\cosh \chi N-\sinh \chi M$, see (\ref{4.10}),  and use (\ref{4.13}), to get
\begin{equation}\label{5.19}
(n\cdot m)={\cosh \chi (N\cdot L)-\sinh \chi \cos\alpha \sin\Theta \over \sqrt{1-\cos^2\alpha \sin^2\Theta}}~~.
\end{equation}
Then (\ref{5.18})  follows from (\ref{5.19}) since $(N\cdot L)\leq (N_s\cdot L)$ for straight MCS in a  flat spacetime \cite{Fursaev:2017aap}.

If the de Sitter universe  is filled in by an homogeneous isotropic background radiation of a certain frequency $\omega$ an MCS generates 
shifts of the energies.  An observer sees the shifts  inside  the string image. This effect is similar to that for MCS in a flat spacetime.
The difference is that the cosmic expansions makes the shifts isotropic at late times, $\delta \omega/\omega
\simeq \delta \omega_2/\omega$.

\section{Summary and Discussion}\label{sum}

The aim of this work was in extending analysis of physical effects caused by MCS in flat spacetimes to MCS moving in an expanding universe.
We generalized the method suggested in \cite{Fursaev:2017aap} to MCS in the de Sitter universe by taking into account isometries of this background. Concrete
results were obtained for flat expanding de Sitter universe, as a most interesting case from the point of view of cosmological applications.

Massless cosmic strings in the de Sitter universe are one-dimensional objects stretched along circles of the Hubble radius. The worldsheet of such MCS
is a 2D null surface on a cosmological horizon. From the point of view of observers  circular MCS move as a result of cosmic expansion.  

Since the holonomies of the MCS background are some isometries of the de Sitter spacetime
the backreaction effects do not change the metric around the string. This situation is analogous to  the  case of massive strings
where a local backreaction is absent  when rotations around the string are isometries.
The effects then are global and result in an angle deficit.

If circular massless cosmic strings appeared at the Planck epoch they would had a microscopic size. The strings then would grow up to
the present Hubble scale. Our description of MCS can be used at quasi de Sitter stages. It is not applicable, however,  to MCS 
during a power low stage,
since the required isometries are absent.  In this case local backreaction effects around MCS should be taken into account.

Circular MCS result in a variety of physical effects. We discussed analogs of the wake and Kaiser-Stebbins effects. An MCS in a universe filled with a dust creates a region of overdensities, which accompanies the string. Overdensities are located outside the string and, thus, have a disc-shaped hole inside. MCS also shift energies of quanta. The shift is composed of (anisotropic) Kaiser-Stebbins-like term and (isotropic) cosmological term. 
Cosmological term dominates at late times. Photons with shifted energies are seen as a spot which fill in the image of the string. 
The both effects  have  features which allow one to distinguish MCS from moving massive cousins. 

A central idea of our approach which allows us to introduce properly the parabolic holonomy around an MCS in locally de Sitter geometry is to
set Cauchy-like data for trajectories on the horizon of the string. This approach can be used to describe another class of physical effects related, for example, to classical
fields on MCS spacetimes. We plan to return to this subject in a separate publication.

\end{document}